# Algorithms for Packet Routing
# in Switching Networks
# with Reconfiguration Overhead

Timotheos Aslanidis[1] and Marios-Evangelos Kogias[2]

[1]National Technical University of Athens, Athens, Greece
taslan.gr@gmail.com

[2]National Technical University of Athens, Athens, Greece
marioskogias@gmail.com

*ABSTRACT*

*Given a set of messages to be transmitted in packages from a set of sending stations to a set of receiving stations, we are required to schedule the packages so as to achieve the minimum possible time from the moment the 1st transmission initiates to the concluding of the last. Preempting packets in order to reroute message remains, as part of some other packet to be transmitted at a later time would be a great means to achieve our goal, if not for the fact that each preemption will come with a reconfiguration cost that will delay our entire effort. The problem has been extensively studied in the past and various algorithms have been proposed to handle many variations of the problem. In this paper we propose an improved algorithm that we call the Split-Graph Algorithm (SGA). To establish its efficiency we compare it, to two of the algorithms developed in the past. These two are the best presented in bibliography so far, one in terms of approximation ratio and one in terms of experimental results.*

## 1. INTRODUCTION

As the need for communication and dissemination of information increases in modern technology based societies, so does the need for faster and more efficient networks and routing of packages between stations. In this context switching networks and the transmission of large packets of data between them, has become an issue of major importance and as information loads continue to increase rapidly, it is expected that the need for well scheduled data transfers to decrease time and resource usage, will keep on being an often addressed subject for many scientists and engineers.

In this manuscript and in the context of message scheduling and transmitting through switching networks we consider the Preemptive Bipartite Scheduling problem (encountered as PBS in bibliography). Given a set of n transmitting stations and a set of m receiving stations, we are required to send across messages, each initiated from a specific transmitter, to reach an also prespecified receiver station. The duration of each message is also predetermined for all messages to be transmitted. Restrictions in the systems considered, are that no transmitter may transmit data towards more than one receiver at any time, nor may a receiver receive, more than one message at a time. Messages are sent in packages and to enhance transmission speed we are allowed to preempt any package and continue transmission of any part of that package at a later time. Unfortunately, since the system has to reconfigure after each preemption, any interruption of packets transmission will come with a time cost. Information on the data that was not sent has to be saved and a new setup has to be initiated for the next packet to start transmitting. Consequently prior to sending any of the packages there will be a setup overhead. We consider this overhead to be constant for all transmission initiations. In this paper we aim to minimize the duration of the aforementioned process.

## 2. PREVIOUS RESULTS

PBS is known to be NP-Complete [9] and proved to be 4/3-ε inapproximable for any ε>0, unless P=NP in [4].

As PBS algorithms can be implemented in various applications, many polynomial time algorithms have been designed to produce solutions close to the optimal, found in [1], [9], [12], [8]. The best guaranteed approximation ratio so far is $2 - \dfrac{1}{d+1}$, where d is the reconfiguration cost, and is found in [1]. Experiments on the performance of various algorithms are presented in [4] and [5].

The problem can be solved in polynomial time if we consider a zero setup cost or if we only want to minimize the number of switchings [9]. Another variation of the problem for which the optimal schedule can be calculated in polynomial time is presented in [1].

For the purposes of this paper we consider 2 algorithms published in the past:

- A-PBS(d+1), found in [1], which so far is the one with the lowest approximation ratio, and

- A1, found in [5], which according to past experiments yields the best experimental results.

To compute each packet to be transmitted, A-PBS(d+1) rounds up the time of each message to the closest multiple of d+1 and calculates the packet reducing the workload of each station to the minimum multiple of d+1.

On the other hand A1 computes an arbitrary packet with a maximum number of messages and decides how to preempt by calculating a lower bound to the remaining transmissions cost to be the minimum possible.

## 3. GRAPH REPRESENTATION AND NOTATIONS

Our data representation will be through a bipartite graph G(V,U,E). V will be the set of transmitters, U the set of receivers while E, the set of edges, will correspond to the messages that have to be transmitted from V to U. A weight (or cost) c(v,u), will be assigned to each of the edges e=(v,u), to denote the time required to transmit the message from node v to node u. Edge weights are considered to be non-negative integers.

Furthermore the following notation will be used: $\Delta=\Delta(G)=\max\{\max_{v\in V}(deg(v)), \max_{u\in U}(deg(u))\}$, that is, $\Delta$ will denote the maximum number of messages that need to be either sent or received from or to any of the stations.

The function t: $V\cup U\rightarrow Z^*_+$ will denote the total workload of any station, namely $t(v)=\sum_{u\in U}c(v,u)$ for any $v\in V$ or $t(u)=\sum_{v\in V}c(v,u)$ for any $u\in U$.

$W=W(G)=\max\{\max_{v\in V}(t(v)), \max_{u\in U}(t(u))\}$, that is W will denote the maximum transmission time of the messages either sent to or received from any station.

$d\in Z^*_+$ will denote the overhead to start the next transmission.

## 4. A HEURISTIC WITH IMPROVED RESULTS

For the purposes of our algorithm the initial graph is split in two parts. $G_M$ comprises edges of weight at least d and $G_m$ contains all edges of weight less than d. Our main concern for $G_M$ is to keep reducing the workload for each of the stations, achieving the minimum transmission time possible, whereas in the case of $G_m$, where edge weights are small in comparison to d, we aim in minimizing the number of switchings. The intuition in designing this algorithm is that for messages of long duration, priority on how to schedule has the message duration rather than the number of preemptions, whilst for messages of shortest duration prioritized is the minimization of the number of preemptions. In particular:

*The Split-Graph Algorithm (SGA)*

Step 1: Split the initial graph G(V,U,E) in two bipartite graphs $G_m(V_m,U_m,E_m)$ and $G_M(V_M,U_M,E_M)$, where $V_m=V_M=V$, $U_m=U_M=U$ and $E_m$ contains all edges of weight less than d, $E_M$ contains all edges of weight d or more. Clearly in this initiation step $E=E_m\cup E_M$ and $E_m\cap E_M=\varnothing$.

Step 2: Use subroutine1 to find a maximal matching M, in $G_M$.

Step 3: Use subroutine 2 to calculate the weight of the matching to be removed. Remove the corresponding parts of the edges.

Step 4: Add edges to M, from $E_m$ to maximize |M| and remove them from $E_m$.

Step 5: Move edges of weight less than d, from the graph induced by step 3 to $E_m$.

Step 6: Repeat steps 2 to 5 until all edges initially in $E_M$ have been completely removed.

Step 7: Use subroutine 3 to calculate $\Delta_m$ maximum matchings in $G_m$, where $\Delta_m$ is the degree of $G_m$.

Step 8: Schedule the messages as calculated in steps 2, 3 and 7.

*Subroutine 1:*

Step 1: M=∅ (Initialization of the matching).

Step 2: For each node $w \in V_M \cup U_M$ calculate t(w).

Step 3: Sort all nodes $w \in V_M \cup U_M$ in decreasing order of t(w). Let L be the induced list of nodes.

Step 4: Let $w_0$ be the 1st node to appear in L. Run sequential search in L to find the 1st neighbor of $w_0$ appearing in L. Denote that neighbor by $w_1$.

Step 5: M←M∪{$w_0$, $w_1$}.

Step 6: Remove $w_0$, $w_1$ from L.

Step 7: Repeat steps 2, to 6 until M becomes maximal.

*Subroutine 2:*

Step 1: For each edge e=(v,u) of the matching M, with corresponding weight c(e) calculate what the value W(G′) of the induced graph G′(V′,U′,E′) would be if all edge weights in the matching were to be reduced by c(e). In this case edges of cost less than c(e) would be completely removed. Set

$$r(e) = \begin{cases} c(e) & , \text{if} \quad W(G') = W(G) - c(e) \\ 0, & \text{otherwise} \end{cases}$$

Step 2: Calculate R=max{r(e) | e∈M}

Step 3: For each edge in M set its new weight c(e)= $\begin{cases} c(e) - R, & \text{if} \quad c(e) > R \\ 0, & \text{otherwise} \end{cases}$ .

*Subroutine 3:*

Step 1: Add nodes and edges to make $G_m$ a regular graph of degree $\Delta_m$. New edges will be of zero weight. In a regular graph, all nodes will be of the same degree.

Step 2: Calculate a maximum matching $M_m$ in $G_m$ and remove all edges of $M_m$ from $G_m$. $G_m$'s degree will now be reduced by 1.

Step 3: Repeat step 2 until $G_m$=∅.

## 5. EXPERIMENTAL RESULTS AND COMPARISON

Figure 1 represents each algorithms' performance in terms of approximation ratio. 1000 test cases have been ran for a 15 transmitters-15 receivers system for values of setup cost varying from 1 to 200 and message durations varying from 1 to 50. SGA performs significantly better than both A1 and A-PBS(d+1) and as the overhead increases it shows an increasingly improved performance. It is important to mention that in practice, as information loads exponentially increase, the number of stations and communication tasks increases and so does the setup cost. That is in fact the most encountered situation nowadays.

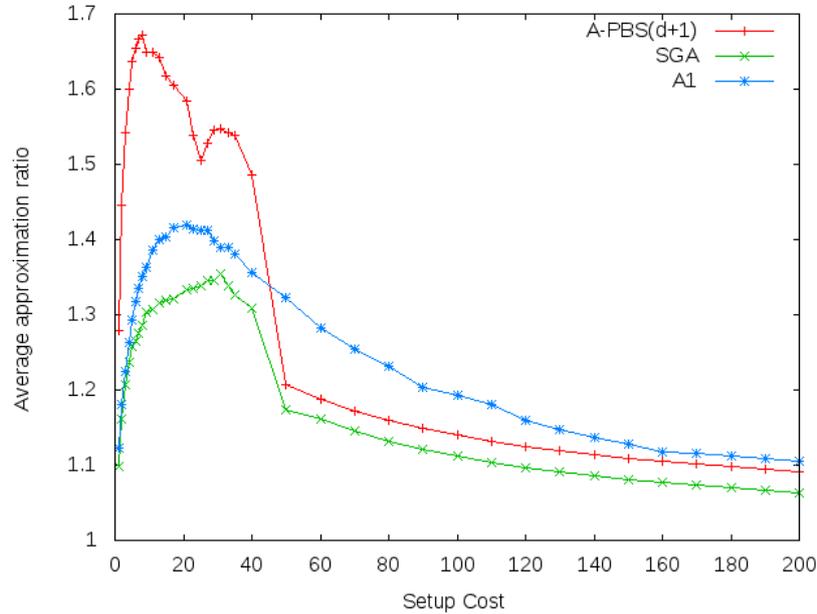

Figure 1. Average approximation ratio comparison

Figure 2 presents the worst performance that each algorithm had depending on the setup cost, in terms of approximation ratio again. SGA is found to be once again a lot more efficient. Furthermore, SGA in most cases has a worst case really close to the average showing that its performance does not fluctuate much, making it an all cases a reliable tool for this type of scheduling.

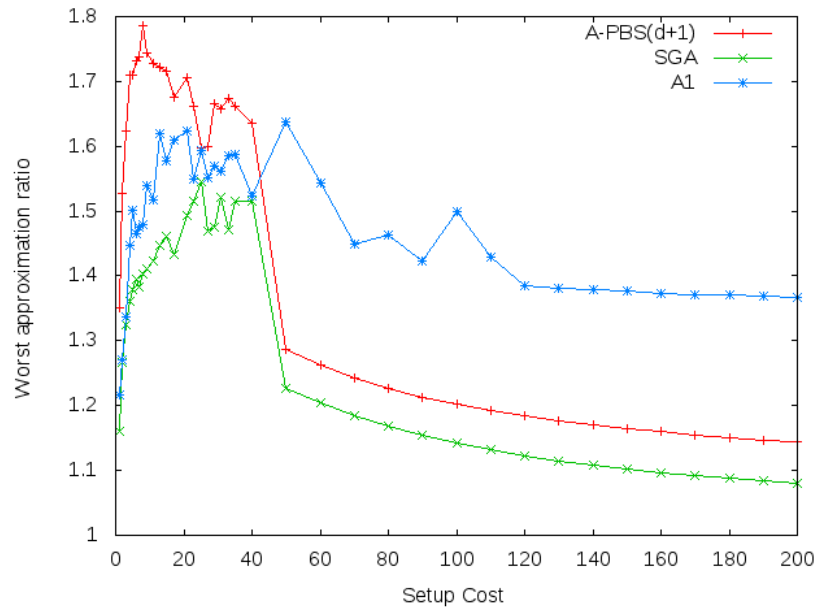

Figure 2. Worst approximation ratio comparison

In terms of running time, SGA also appears to be a lot more efficient, as experiments have shown that A1 and A-PBS(d+1) are by up to 500% slower. This is mainly because SGA is based on an entire different approach on how to schedule the messages, using subroutines that in general are a lot faster than those used in previous papers.

# 6. DISCUSSION AND FUTURE WORK

Our newly presented algorithm (SGA), has proven to produce much more efficient routings both in terms of hardware usage as well as time span. Therefore, we believe that the idea of splitting the initial graph in parts can be further researched and depending on the magnitude of the edges as well as the setup cost, the Split-Graph Algorithm's efficiency can be further improved. An approximation ratio for SGA could be established to be less than 2. Exploiting to a greater extend algorithms that provide optimal solutions for special instances of the problem might also yield interesting new approximation algorithms. Finally, lifting limitations of the problem or introducing new ones could help in developing new classes of graphs for which polynomial algorithms might provide an optimal schedule. Such algorithms could be the tools to designing new and improved approximation algorithms.

# REFERENCES

[1] F. Afrati, T. Aslanidis, E. Bampis, I. Milis, Scheduling in Switching Networks with Set-up Delays. Journal of Combinatorial Optimization, vol. 9, issue 1, p.49-57, Feb 2005.

[2] G. Bongiovanni, D. Coppersmith and C. K.Wong, An optimal time slot assignment for an SS/TDMA system with variable number of transponders, IEEE Trans. Commun. vol. 29, p. 721-726, 1981.

[3] J. Cohen, E. Jeannot, N. Padoy and F. Wagner, Messages Scheduling for Parallel Data Redistribution between Clusters, IEEE Transactions on Parallel and Distributed Systems, vol. 17, Number 10, p. 1163, 2006.

[4] J. Cohen, E. Jeannot, N. Padoy, Parallel Data Redistribution Over a Backbone, Technical Report RR-4725, INRIA-Lorraine, February 2003.

[5] P. Crescenzi, X. Deng, C. H. Papadimitriou, On approximating a scheduling problem, Journal of Combinatorial Optimization, vol. 5, p. 287-297, 2001.

[6] R. L. Cruz and S. Al-Harthi, Packet Scheduling with Switch Configuration Delays, in Proc. 39th Annu. Allerton Conf. Commun., Contr., Comput., 2001.

[7] S. Even, A. Itai, A. Shamir, On the complexity of timetable and multicommodity flow problems SIAM J. Comput., vol. 5, p. 691-703, 1976.

[8] I. S. Gopal, G. Bongiovanni, M. A. Bonucelli, D. T. Tang, C. K. Wong, An optimal switching algorithm for multibeam satellite systems with variable bandwidth beams, IEEE Trans. Commun. vol. 30, p. 2475-2481, Nov. 1982.

[9] I. S. Gopal, C. K. Wong Minimizing the number of switchings in an SS/TDMA system IEEE Trans. Commun. vol. 33, p. 497-501, 1985.

[10] T. Inukai, An efficient SS/TDMA time slot assignment algorithm IEEE Trans. Commun. vol 27, p. 1449-1455, Oct. 1979.

[11] E. Jeannot and F. Wagner, Two fast and efficient message scheduling algorithms for data redistribution over a backbone, 18th International Parallel and Distributed Processing Symposium, 2004.

[12]    A. Kesselman and K. Kogan, Nonpreemptive Scheduling of Optical Switches, IEEE Transactions in Communications, vol. 55, number 6, p. 1212, 2007.

[13]    K. S. Natarajan and S. B. Calo, Time slot assignment in an SS/TDMA system with minimum switchings IBM Res. Rep. 1981.

[14]    B. Towles and W. J. Dally, Guaranteed Scheduling of Switches with Configuration Overhead, in Proc. Twenty-First Annual Joint Conference of the IEEE Computer and Communications Societies INFOCOM '02. pp. 342-351, June 2002.

**Authors**

Timotheos Aslanidis was born in Athens, Greece in 1974. He received his Mathematics degree from the University of Athens in 1997 and a master's degree in computer science in 2001. He is currently a phd candidate at the National and Technical University of Athens in the department of electrical and computer engineering. His research interests comprise but are not limited to computer theory, number theory, network algorithms and data mining algorithms.

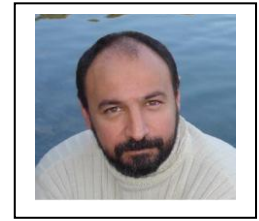

Marios-Evangelos Kogias was born in Athens in 1991. Currently, he is an undergraduate student at the National Technical University of Athens pursuing a diploma in Electrical and Computer Engineering. He is a passionate coder and interested mainly in operating systems, computer networks and algorithms.

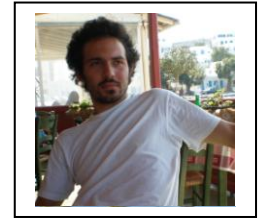